# Plasma polarization in massive astrophysical objects

I.L. Iosilevskiy

*JIHT RAS, MIPT (State University), Moscow, Russia*

**Absract.** Macroscopic plasma polarization, which is created by gravitation and other mass-acting (inertial) forces in massive astrophysical objects (MAO) is under discussion. Non-ideality effect due to strong Coulomb interaction of charged particles is introduced into consideration as a new source of such polarization. Simplified situation of totally equilibrium isothermal star without relativistic effects and influence of magnetic field is considered. The study is based on density functional approach combined with "local density approximation". It leads to conditions of constancy for generalized (electro)chemical potentials and/or conditions of equilibrium for the forces acting on each charged specie. New "non-ideality force" appears in this consideration. Hypothetical sequences of gravitational, inertial and non-ideality polarization on thermo- and hydrodynamics of MAO are under discussion.

## 1. Introduction

Long-range nature of Coulomb and gravitational interactions leads to specific manifestation of their joint action in massive astrophysical objects (MAO). The main of them is polarization of plasmas under gravitational attraction of ions. Extraordinary smallness of gravitational field in comparison with electric one (the ratio of gravitational to electric forces for two protons is ~ $10^{-36}$) leads to the fact that extremely small and thermodynamically (energetically) negligible deviation from electroneutrality can provide thermodynamically noticeable (even significant) consequences at the level of first (thermodynamic) derivatives. This is the main topic of present paper.

## 2. Electrostatics of massive astrophysical objects.

Gravitational attraction polarizes plasma of massive astrophysical bodies due to two factors: (*i*) smallness of electronic mass in comparison with ionic one and (*ii*) general non-uniformity of MAO due to long-range nature of gravitational forces. The first, mass-dependent type of gravitational polarization is part of more general phenomenon: (**A**) - any inertial (mass-acting) force (due to rotation, vibration, inertial expansion and compression *etc.*) polarizes ion-electron plasma due to the same reason: low mass of electron in comparison with that of ions. The second type of discussed polarization is also part of more general phenomenon (see for example [1]): (**B**) - any non-uniformity in equilibrium Coulomb system is accompanied by its polarization and existence of stationary profile of average electrostatic potential. Important particular case is existence of stationary drop of average electrostatic potential (*Galvani potential*) at any two-phase interface in equilibrium Coulomb system [2] (see also [1-10]).

Plasma polarization at *micro*-level is well known in classical case as Debye-Hueckel screening [15] and in the case of degenerated electrons as Thomas-Fermi screening [16]. Plasma polarization under gravitational forces at *macro*-level (*macroscopic screening*) is less known although it was claimed [17] and proved at the same years [11][12].

Remarkable feature of gravitational polarization is that resulting average electrostatic field *must be* of the *same order* as gravitational field (counting per one proton). Average electrostatic force $F_E^{(p)}$ must be equal to *one half* of gravitational force $F_G^{(p)}$ in ideal, isothermal and non-degenerated electron-proton plasma of outer layers of a star [11][12]. This *compensation* is supposed to be equal just *twice* gravitational force (counting per one proton) in the case of ionic plasma on strongly *degenerated* electronic *background* in compact stars (white dwarfs, neutron stars etc) [13][5]. It seems natural to suppose that in general case of electron-ionic system the value of discussed compensation (counting per one ion $Z$) lay between these two limits:

$$\text{(at } n_e\lambda^3_e << 1) \quad -[Z/(1+Z)] \geq F_E^{(Z)}/F_G^{(Z)} \geq -1 \quad \text{(at } n_e\lambda^3_e >> 1) \tag{1}$$

It was claimed [18][19]that inequality (1) is valid for ideal-gas assumption *only* (with arbitrary degree of electron degeneracy). If one takes into account non-ideality effects, there may be conditions when polarization force *overcompensates* gravitational force i.e. $|F_E^{(Z)}| \geq |F_G^{(Z)}|$ due to additional non-ideality effect (see below).

Real plasmas of compact stars (white dwarfs and neutron stars) are close to isothermal conditions due to high thermal conductivity of degenerated electrons. At the same time plasma of ordinary stars, for example, of the Sun, is not isothermal. Temperature profile, heat transfer and thermo-diffusion exist in such plasmas. It should be taken into account self-consistently in calculation of average electrostatic field.

## 3. Gravitational polarization with non-ideality effects.

Let's consider simplified case of hypothetical totally equilibrium non-uniform self-gravitating body without taking into account relativistic effects and influence of magnetic field. Approach accepted in [13][5][20] etc. operates with idea of individual *partial pressures* for electrons and ions, $P_e(n_e,T)$ and $P_i(n_i,T)$, and based on solution of several separate ("*partial*") *hydrostatic equilibrium* equations for each specie of particles instead of standard *unique* hydrostatic equilibrium equation for *total pressure* and total mass density [21][22]. It should be stressed [18][19] that partial pressures and partial hydrostatic equilibrium equations are not well-defined quantities in general case of equilibrium non-ideal system [1]. General approach for description of thermodynamic equilibrium in this case is multi-component variational formulation of statistical mechanics [23][24][25]. Thermodynamic equilibrium conditions may be written in three forms: (*i*) – extremum condition for thermodynamic potential of total system (free energy functional) regarding to variations of one-, two-, three-particle etc. correlations in the system; (*ii*) – constancy conditions for generalized "electro-chemical" potentials [2] for all species (electrons, ions etc.), and (*iii*) – zero conditions for the sum of (generalized) average forces acting on each specie of particles in the system. The problem is that all these quantities: (*i*) total thermodynamic potential and partial (*ii*) electrochemical potentials and (*iii*) average forces are essentially non-local functionals on mean-particle correlations. Standard technique is separation of main non-local parts of free energy functional – electrostatic and gravitational energies in mean-field approximation [26]. It is *assumed* that all non-local effects are *exhausted* by these two terms. Consequently next standard technique is the "local-density" approximation for the rest free energy term $F^*[n_i(\mathbf{r}), n_e(\mathbf{r}), T]$ of hypothetical non-ideal charge system with extracted electrostatic and gravitational energies in mean-field approximation. It should be stressed [27][28] that (local) free energy density $f^*(n_i, n_k, ..., T)$ must be defined as thermodynamic limit of specific free energy of (new) *uniform* macroscopic *non-electroneutral* multi-component charge system with charge particle densities $(n_j, n_k, ...)$ on *compensating Coulomb (and strictly speaking gravitational) background(s)* [1][27]. It leads to two sets of local forms for thermodynamic equilibrium condition: in terms of electrochemical potentials and generalized thermodynamic forces (2, 3):

$$m_j \varphi_G(\mathbf{r}) + q_j \varphi_E(\mathbf{r}) + \mu_j^{(chem)}\{n_i(\mathbf{r}), n_e(\mathbf{r}); T\} = \mu_j^{(el.chem)} = \text{const} \qquad (j = \text{electrons, ions}) \quad (2)$$

$$m_j \nabla\varphi_G(\mathbf{r}) + q_j \nabla\varphi_E(\mathbf{r}) + \nabla\mu_j^{(chem)}\{n_i(\mathbf{r}), n_e(\mathbf{r}); T\} = \nabla\mu_j^{(el.chem)} = 0 \qquad (j = \text{electrons, ions}) \quad (3)$$

Here $\varphi_G(\mathbf{r})$ and $\varphi_E(\mathbf{r})$ – gravitational and electrostatic potentials, $\mu_j^{(chem)}$ and $\mu_j^{(el.chem)}$ – local chemical and non-local electrochemical potentials, $m_j$ and $q_j$ – mass and charge of specie $j$ ($q_j \equiv Z_j e$; $j$ = i,e), $\nabla\varphi(\mathbf{r})$, $\nabla\mu(\mathbf{r})$ – spatial gradients. It should be stressed that the set of equations (5) and (6) are well-defined equivalents (substitute) for the set of separate equations of hydrostatic equilibrium for mentioned above partial pressures and densities of charged species in ideal-gas conditions [11][12].

The smallness of gravitational forces in comparison with Coulomb ones leads to the fact that the electroneutrality conditions are valid for almost everywhere in MAO with precision $\sim 10^{-36}$. At the same time it should be stressed that in finite number of thin layers, for example on phase transition interfaces etc., electroneutrality conditions may be violated. This is of primary importance for hydrodynamic and diffusion processes in the vicinity of such interfaces (see below). As for the dominating electroneutral zones of stationary or rotating star, compendious form of equilibrium conditions could be derived for general case of non-ideal multi-component charge system with arbitrary degree of electronic degeneracy (without relativistic effects and magnetic field):

$$e\nabla \varphi_E(\mathbf{r}) = -\nabla \varphi_M(\mathbf{r}) \frac{\langle Z | \mathbf{D}_\mu^n | M \rangle}{\langle Z | \mathbf{D}_\mu^n | Z \rangle}, \qquad \text{here } \langle A | \equiv | A \rangle \equiv \{a_j\} \text{ and } \mathbf{A} \equiv \| a_{jk} \|, \qquad (4)$$

Here $\varphi_E(\mathbf{r})$ is charge-acting, and $\varphi_M(\mathbf{r}) \equiv \varphi_G(\mathbf{r}) + \varphi_R(\mathbf{r})$ is mass-acting potential – the sum of gravitational and centrifugal ones. $\langle Z | \equiv \{Z_j\}$ and $| M \rangle \equiv \{M_j\}$ are the charge and mass vectors and $\mathbf{D}_\mu^n(\mathbf{r})$ is Jacobi matrix $(\partial \mathbf{n}/\partial \mathbf{\mu})_T \equiv \|(\partial n_j/\partial \mu_k)_T\|$ ($j,k = 1,2,3,\ldots$).

**4. "Physical" and "chemical" representations**.
The problem for application of formula (4) in real astrophysical situations is adequate choice of basic varying quantities $n_j(\mathbf{r})$ in free energy functional (well-known dilemma of "physical" and "chemical pictures"). This problem is similar, but more complicated than that in traditional theory of non-ideal plasmas [1]. The point is that the same representation may be "physical" or "chemical" depending on application. For example the set of basic variables as hydrogen and helium nuclei and electrons, is "physical" one for description of Jupiter, Saturn and extrasolar planets interior. At the same time this choice of basic variables is "chemical" one for the case of compact stars interiors, where nuclear and $\beta$-decay reactions between free and bound neutrons and protons are equilibrium and adequate "physical" representation corresponds to the choice of protons and electrons as independent variables [22]. Again, the same choice (protons and electrons) will correspond to "chemical picture" in the case of interiors of so-called *hybrid stars* – combination of quark matter core with hadronic crust. The proper choice of basic variables for "physical" representation in this case corresponds to combination of one quark specie (*u, d, s*) and electrons [22] [35]. It should be stressed that in any variant of mentioned above "chemical" representation minimization of free energy functional under condition of conservation of electric and baryon charge leads to *generalized* form of equations for chemical, ionization and other type equilibrium reactions (*ab = a + b*)

$$\mu_{ab}(\mathbf{r}) = \mu_a(\mathbf{r}) + \mu_b(\mathbf{r}) \qquad (5)$$

Equatios (3)-(5) are valid both variant: "physical" and "chemical" pictures within proper choice of vectors $\langle Z(\mathbf{r}) |$ and $| M(\mathbf{r}) \rangle$ and matrix $\mathbf{D}_\mu^n(\mathbf{r})$ with corresponding non-ideality corrections for mutual effective interactions of "free" particles.

**5. Chemical picture. Ideal mixture approximation**.
This approximation is useful for application in outer layers of MAO when weakly degenerated electrons are partially localizes on nuclei with creating of charged (ions) and neutral complexes (atoms, molecules, clusters etc.). In hypothetical isothermal conditions the mater is described approximately as ideal multi-component mixture of mutual (arbitrary) species under the local chemical and ionization equilibrium (5). Matrix $\mathbf{D}_\mu^n(\mathbf{r})$ in ideal-mixture approximation became diagonal. The equation (4) is simplified in this approximation.

$$e\nabla\varphi_E(\mathbf{r}) = -\nabla\varphi_M(\mathbf{r})\left(\sum_j \tilde{n}_j M_j Z_j\right)\left(\sum_j \tilde{n}_j Z_j^2\right)^{-1}, \qquad \tilde{n}_j \equiv kT\left(\partial n_j/\partial\mu_j\right)^{id.gas}_{T,n_{k\neq j}} \quad (j=1,2,3\ldots) \quad (6)$$

Note that electronic contribution falls out from (6) in the limit of strong electron degeneracy due to diminishing of ideal-gas electronic compressibility: $\tilde{n}_e \to 0$ $(n_e\lambda_e^3 \gg 1)$. In the limit of ideal non-degenerated two-component electron-ionic plasma equation (6) coincides with (1).

**6. Comments. Thermodynamics**:
- In contrast to the ideal-gas approximation (6) equation (4) in general case describes equilibrium conditions as competition between not *two*, but *three* sources of influence: gravitation field, polarization field and generalized "non-ideality force".
- Coulomb "non-ideality force", when it is taken into account in (4), moves positive ions *inside* the star in addition to gravitation. Hence "non-ideality force" *increases* compensating electrostatic field $\nabla\varphi_E(\mathbf{r})$ in comparison with ideal-gas approximation (6).
- In the case of classical (non-degenerated) plasma the matrix $\mathbf{D}^n_\mu(\mathbf{r})$ in (4) depends on Coulomb non-ideality. In the weak non-ideality limit for two-component electron-ionic ($Z$) plasma it could be described in Debye-Hueckel approximation.
- The non-ideality matrix $\mathbf{D}^n_\mu(\mathbf{r})$ in (4) may be *negative* and the multiplyer in right side of (4) may be *less* than *unity* in the case when strongly non-ideal ionic subsystem is combined with highly degenerated and almost ideal electrons (for example, in white dwarfs). In this case one meets "*overcompensation*" when polarization field could be *higher* (by absolute value) than gravitation field: i.e. $|F_E^{(Z)}| > |F_G^{(Z)}|$. Rough and accurate approximations for EOS of OCP($Z$) could be found elsewhere (see for example [30,31]).
- Any jump-like discontinuity in local thermodynamic state, in particular, phase transition interface or the set of interfaces between *mono-ionic layers* with different masses $M_i$, charges $Z_i$ and non-ideality parameters $\Gamma_Z$ in neutron star crust [32], leads in general case to corresponding jump-like discontinuity in "non-ideality force" in (4) and consequently, to jump-like discontinuity in final polarization field $\nabla\varphi_E(\mathbf{r})$. It means in its turn appearance of *macroscopic charge* at all discussed inter-phase and inter-layer *interfaces* in addition to electrostatic potential drop (Galvani potential) mentioned at the beginning of the paper.
- Equation (4) is not restricted by spherical symmetry conditions. It is valid for rotating stars and stars in binary systems etc. It is valid for any self-gravitating system in *total thermodynamic equilibrium* (see above comment about non-isothermal state)

**Comment. Hydrodynamics**.
Plasma polarization could suppress hydrodynamic instabilities in MAO. For example, plasma polarization could suppress hypothetical Rayleigh-Taylor instability in liquid mixture of nuclei $\{_{16}O^{8+}, _{12}C^{6+}, _4He^{2+}\}$ in interior of typical white dwarf in the vicinity of its freezing boundary [33]. Accordingly (4) polarization field compensates almost totally gravitation field acting on any nucleus, O, C and He, due to their symmetry ($A/Z = 2$) so that the total force is roughly equal to zero: $(F_E^{(Z)} + F_G^{(Z)} \approx 0)$. The final weak discrimination in total force acting on each ion in the mixture $\{_{16}O^{8+} + _{12}C^{6+} + _4He^{2+}\}$ depends on interplay between Coulomb non-ideality effects and electron degeneracy [29].

**Acknowledgments**
The work was supported by Grants: ISTC 3755 and by RAS Scientific Programs "Research of matter under extreme conditions" and "Physics of high pressure and interiors of planets"